\documentclass[twocolumn,showpacs,preprintnumbers,amsmath,amssymb]{revtex4}

\def\undersim#1{\setbox9\hbox{${#1}$}{#1}\kern-\wd9\lower
    2.5pt \hbox{\lower\dp9\hbox to \wd9{\hss $_\sim$\hss}}}
\usepackage{amssymb}
\usepackage{amsmath}
\usepackage{graphicx}

\def\undersim#1{\setbox9\hbox{${#1}$}{#1}\kern-\wd9\lower
    2.5pt \hbox{\lower\dp9\hbox to \wd9{\hss $_\sim$\hss}}}

\def\mk{{\mathbf k}}

\begin{document}

\title{Effect of in-medium mass-shift on transverse-momentum spectrum and elliptic anisotropy of $\phi$ meson}

\author{Yong Zhang$^{1,\,2}$\footnote{zhy913@jsut.edu.cn}}
\author{Jing Yang$^{3,}$}
\author{Weihua Wu$^{1}$}
\affiliation{\small$^1$School of Mathematics and Physics,
Jiangsu University of Technology, Changzhou, Jiangsu 213001, China\\
$^2$School of Science,Inner Mongolia University of Science $\&$ Technology, Baotou,Inner Mongolia Autonomous Region 014010, China\\
$^3$School of Physics and School of International Education Teachers, Changchun Normal University, Changchun, Jilin 130032, China
   }

\date{\today}

\begin{abstract}
We study the effect of in-medium mass-shift on transverse-momentum spectrum and elliptic anisotropy of $\phi$ meson.
It is found that the mass-shift enhances the $\phi$ yields and suppresses the elliptic flow $v_2$ in large momentum region, and the effects increase with the increasing mass-shift. The effects are various for different sources and decrease with the increasing expanding velocity. We further study the effects for parts of all $\phi$ meson with mass-shift, and the effects decrease with the decreasing probability of $\phi$ meson with mass-shift.
Since the different mass-shift lead to different transverse-momentum spectrum and $v_2$, our study may provide a
way to restrict the ranges of in-medium mass-shift of $\phi$ meson in high-energy heavy-ion collisions.


\end{abstract}

\maketitle

\section{Introduction}
The main goal of high-energy heavy-ion collisions is to create the matter of
quark-gluon plasma (QGP), and study its properties by analyzing the observables of particles in final
state \cite{BRAHMS_NPA2005,PHOBOS_NPA2005,STAR_NPA2005,PHENIX_NPA2005}. $\phi$(s$\bar{s}$) meson is
an excellent probe for studying QGP because it is sensitive to several aspects of the collision
\cite{Shor_PRL1985,PHENIX_PRC2005,STAR_PRL2007,PHENIX_PRL2007,STAR_PLB2009,PHENIX_PRC2011,PHENIX_PRC2015}
and it is expected to have a small interaction with the hadronic medium \cite{Shor_PRL1985,STAR_PRL2007,PHENIX_PRL2007,JHChen_PRC06,Hirano_PRC08,Nasim_PRC13,STAR_PRL16}.
However, the recent  experimental data of the elliptic flow of identified hadrons in the Pb-Pb collisions at
$\sqrt{s_{NN}}=2.76$ GeV indicate that the $\phi$ meson may have a larger hadronic cross section than its
current theoretical estimate \cite{ALICE_JHEP15}. It is still an open issue to determine the interaction
between $\phi$ and the hadronic medium.

The transverse-momentum spectrum and elliptic flow of $\phi$ meson are
important observables in high-energy heavy-ion collisions \cite{PHENIX_PRC2005,STAR_PRL2007,PHENIX_PRL2007,STAR_PLB2009,PHENIX_PRC2011,PHENIX_PRC2015,STAR_PRL16,ALICE_JHEP15}. The transverse-momentum spectrum can reveal information about the thermalization and expansion
of the $\phi$ emission source \cite{PHENIX_PRC2005,STAR_PLB2009,PHENIX_PRC2011,PHENIX_PRC2015},
and the elliptic flow $v_2$ of $\phi$ meson was an important evidence for the formation
of hot and dense matter with partonic collectivity \cite{STAR_PRL2007,PHENIX_PRL2007,STAR_PRL16,ALICE_JHEP15}.
The $\phi$ meson is expected to have a mass-shift in the hadronic medium, and the mass-shift are various for several theoretical calculations \cite{Hatsuda,ask_npa94,Song_PLB,Smith_PRC98,Lin_npa02,Fuch_epja,Cab_PRC03,Metag}.
In this work, we will study the effect of in-medium mass-shift on transverse-momentum spectrum and elliptic flow $v_2$
of $\phi$ meson, by using the ideal relativistic hydrodynamics in $2 + 1$ dimensions to describe the
transverse expansion of sources with zero net baryon density and combine the Bjorken boost-invariant
hypothesis \cite{Bjorken_PRD83} for the source longitudinal evolution. In the calculations, we use
the equation of state of s95p-PCE \cite{Shen_PRC10} and take the initial energy density distribution
in the transverse plane as the Gaussian distribution: $\epsilon = \epsilon_0\exp[-x^2/(2R_x^2)-y^2/(2R_y^2)]$,
where $\epsilon_0$ and $R_i~(i=x,y)$ are the parameters of the initial source energy density and radii \cite{YZHANG_PRC15}. And the freeze-out temperature of $\phi$ is taken as $140$ MeV \cite{AsaCsoGyu99,Padula06,YZHANG16,YZHANG_PRC15}.

This paper is organized as follows. In Sec.II, we present the formulas of the single-particle
momentum distribution for boson with in-medium mass-shift. In Sec. III, the effect
of mass-shift on the transverse momentum spectrum and the elliptic flow $v_2$ of $\phi$
meson will be shown. Finally, a summary of this paper are given in Sec. IV.

\section{Formulism}
Denote $a_\mk\, (a^\dagger_\mk)$ the annihilation (creation) operator of the free boson
with momentum $\mk$ and mass $m$, and the invariant single-particle momentum distribution
can be expressed by
\begin{equation}
N(\mk)=\omega_\mk\frac{d^{3}N}{d\mk}=\omega_\mk\langle a^\dagger_{\mk} a_{\mk} \rangle,
\end{equation}
where $\langle \cdots \rangle$ means the thermal average, and $\omega_\mk=\sqrt{\mk^2 + m^2}$
is the energy of the free particle. For a homogeneous source with volume $V$ and temperature
$T$, the single particle momentum distribution becomes
\begin{equation}
N(\mk)=\frac{V}{2\pi^3}\omega_\mk n_B(\mk),
\end{equation}
\begin{equation}
n_B(\mk)=\frac{1}{\exp(\omega_{\mk}/T)-1}.
\end{equation}
Denote $b_\mk\, (b^\dagger_\mk)$ the annihilation
(creation) operator of the boson with momentum $\mk$ and modified mass $m_{\!*}$ ($m_{\!*}=m-\delta \!m$) in
hadronic medium. The operators $(a_\mk\,,a^\dagger_\mk)$ and $(b_\mk\,,b^\dagger_\mk)$
were related by the Bogoliubov transformation \cite{AsaCso96,AsaCsoGyu99}
\begin{equation}
a_{\mk} = c_{\mk}\,b_{\mk} + s^*_{-\mk}\,b^\dagger_{-\mk}.
\vspace*{-2mm}
\end{equation}
For a homogeneous source, $c_{\mk} = \cosh f_{\mk}$, $s_{\mk} = \sinh f_{\mk}$ and
$f_{\mk} = \frac{1}{2} \log (\omega_{\mk}/\Omega_{\mk})$. Where $\Omega_\mk=\sqrt{\mk^2+m_{\!*}^2}$
is the energy of the particle in medium, and the single particle momentum distribution becomes \cite{AsaCsoGyu99,Padula06}
\begin{equation}
N(\mk)=\frac{V}{2\pi^3}\omega_{\mk}n_1(\mk),
\end{equation}
\begin{equation}
n_1(\mk)=|c_{\mk}|^2\,n_{\mk}+|s_{-\mk}|^2(n_{-\mk}+1),
\end{equation}
\begin{equation}
n_{\mk}=\frac{1}{\exp(\Omega_{\mk}/T)-1},
\end{equation}
as can be seen from equation $(5)-(7)$, the single particle momentum distribution $N(\mk)$
changes with the particle mass in medium.

For hydrodynamic sources, the single particle momentum distribution of the boson without in-medium mass-shift
was expressed as \cite{Cooper,KolHei03,Kol00}
\begin{eqnarray}\label{SP0}
 N_0(\mk)\!=\!\int \frac{g_i}{(2\pi)^3}d^4\sigma_{\mu}(r)k^\mu\,n^0_{\mk'},
\end{eqnarray}
\begin{equation}\label{BZ0}
n^0_{\mk'}=\frac{1}{\exp(\omega_{\mk'}(r)/T(r))-1},
\end{equation}
\begin{eqnarray}
\omega'_{\mk'}(r)=\sqrt{\mk'^2(r)+m^2}=k^{\mu} u_{\mu}(r),
\end{eqnarray}
and the single particle momentum distribution of the boson with in-medium mass-shift
was expressed as \cite{AsaCsoGyu99,Padula06,YZHANG_PRC15,YZHANG_CPC15}
\begin{eqnarray}\label{SP}
 N_*(\mk)\!&&=\!\int \frac{g_i}{(2\pi)^3}d^4\sigma_{\mu}(r)k^\mu\, \! \Bigl\{|c'_{\mk'}|^2\,
n'_{\mk'}\nonumber\\
&&\hspace*{4.2mm}+\,|s'_{-\mk'}|^2\,[\,n'_{-\mk'}+1]\Bigr\},
\end{eqnarray}
\begin{equation}\label{csk}
c'_{\pm\mk'}=\cosh[\,f'_{\mk'}\,], \,\,\,s'_{\pm\mk'}=\sinh[\,f'_{\mk'}\,],
\end{equation}
\begin{eqnarray}
f'_{\mk'}=\frac{1}{2} \log \left[\omega'_{\mk'}/\Omega'_{\mk'}\right]
=\frac{1}{2}\log\left[k^{\mu}u_{\mu}(r)/k^{*\nu}u_{\nu}(r)
\right],
\end{eqnarray}
\begin{eqnarray}
&&\hspace*{-7mm}\Omega'_{\mk'}(r)=\sqrt{\mk'^2(r)+m_*^2}\nonumber\\
&&\hspace*{3.9mm}=\sqrt{[k^{\mu} u_{\mu}(r)]^2-m^2+m_*^2}\nonumber\\
&&\hspace*{3.8mm}=k^{*\mu} u_{\mu}(r),
\end{eqnarray}
\begin{equation}\label{BZ}
n'_{\pm\mk'}=\frac{1}{\exp(\Omega_{\mk'}(r)/T(r))-1},
\end{equation}
where $g_i$ is the degeneracy factor for hadron species $i$. The quantity
$d^4\sigma_{\mu}(r)$ is the four-dimension element of freeze-out hypersurface,
and $u_{\mu}(r)$, $T(r)$ are the source four-velocity and temperature at freeze-out, respectively.
$k^{\mu}=(\omega_{\mk},{\mk})$ is the four-momentum of the particle, and ${\mk}'$
is the local-frame momentum corresponding to $\mk$. If there is no
mass-shift, the coefficients $|c'_{\mk'}|^2$ and $|s'_{\mk'}|^2$ will be 1 and 0, respectively.
And $N_*(\mk)$ will be equal to $N_0(\mk)$.

\section{Results}
In this section, we will present the results in three subsections. In section A,
we will show the effect of in-medium mass-shift on the transverse momentum spectrum of $\phi$.
In section B, the effect of in-medium mass-shift on the elliptic flow $v_2$ will be shown. We further study
the effect for parts of all $\phi$ mesons with mass-shift in section C.
\subsection{Effect of in-medium mass-shift on the transverse momentum spectrum}
\begin{figure}[htbp]
\includegraphics[scale=0.51]{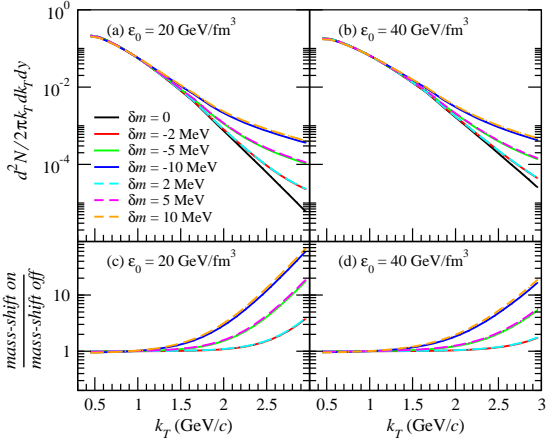}
\vspace*{0mm}
\caption{(Color online) The normalized transverse momentum spectrum of $\phi$ meson with various mass-shift (top panel)
 and the ratio of the normalized transverse momentum spectrum with mass-shift to which without mass-shift
(bottom panel) for the initial conditions $\epsilon_0=$ 20 and 40 GeV/fm$^3$ and $R_x=3$ fm, $R_y=4$ fm. }
\label{psp}
\end{figure}

In Fig. \ref{psp} (a) and (b), we show the normalized transverse momentum spectrum of $\phi$
meson with various mass-shift $\delta$m for the initial conditions $\epsilon_0=$ 20 and 40 GeV/fm$^3$
and $R_x=3$ fm, $R_y=4$ fm. The ratio of the normalized transverse momentum
spectrum with mass-shift to which without mass-shift is also shown in Fig. \ref{psp} (c) and (d).
A small mass-shift leads to an increase in the yield of $\phi$ meson in large transverse momentum
regions and thus reduce the slope of the transverse momentum spectrum, and the effect increases with
the increasing mass-shift. For a fixed mass-shift, the transverse momentum spectrum for the initial
conditions $\epsilon_0=$ 20 GeV/fm$^3$ is more affected than for $\epsilon_0=$ 40 GeV/fm$^3$.

With the Eq.\,(\ref{csk}) and (\ref{BZ}), the Eq.\,(\ref{SP}) can be rewritten as
\begin{eqnarray}\label{SP1}
N_*(\mk)\!=\!\int \frac{g_i}{(2\pi)^3}d^4\sigma_{\mu}(r)k^\mu\, \! [F_1\,
n'_{\mk'}+F_2],
\end{eqnarray}
\begin{equation}
F_1=1+2F_2,\,\,\,F_2=|s'_{\mk'}|^2,
\end{equation}
the quantity $n'_{\mk'}$ in Eq.\,(\ref{SP1}) is approximately equal to the quantity
$n^0_{\mk'}$ in Eq.\,(\ref{SP0}) for a small mass-shift, so the coefficient $F_2$
is the main factor for the particle spectrum.

\begin{figure}[htbp]
\includegraphics[scale=0.51]{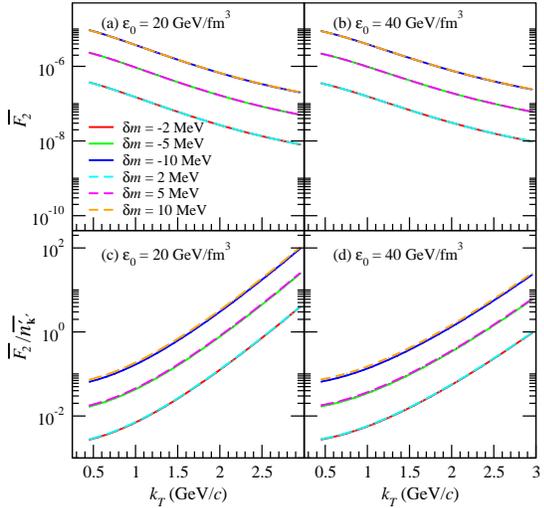}
\vspace*{0mm}
\caption{(Color online) The average coefficient $\overline{F_2}$ with various mass-shift (top panel) and the ratio of $\overline{F_2}$ to $\overline{n'_{\mk'}}$, where "---" means the average over all freeze-out points.}
\label{psk}
\end{figure}

In Fig. \ref{psk}, we show the average coefficients $\overline{F_2}$ with various mass-shift and
the ratio of $\overline{F_2}$ to $\overline{n'_{\mk'}}$, where "---" means the average over all
freeze-out points. The average coefficient $\overline{F_2}$ is very small (see Fig. \ref{psk} (a) and (b))
and the coefficient $F_2$ is positive, so the coefficient $F_1$ is approximately equal to 1 and the Eq.\,(\ref{SP1})
can be rewritten as
\begin{eqnarray}\label{SP2}
N_*(\mk)\!=\!\int \frac{g_i}{(2\pi)^3}d^4\sigma_{\mu}(r)k^\mu\, \! [\,
n'_{\mk'}+F_2].
\end{eqnarray}
The ratio of $\overline{F_2}$ to $\overline{n'_{\mk'}}$ increases with the increasing transverse momentum,
so the effect of in-medium mass-shift on spectrum increases with the increasing transverse momentum (see Fig. \ref{psp}). With the same mass-shift, the average coefficients $\overline{F_2}$ are almost the same for the initial conditions $\epsilon_0=$ 20 and 40 GeV/fm$^3$. The high initial energy density leads to a large expanding velocity and
thus increases the yield ${n'_{\mk'}}$ for large transverse momentum, so the ratio of $\overline{F_2}$ to $\overline{n'_{\mk'}}$ for $\epsilon_0=$ 40 GeV/fm$^3$ is smaller than for $\epsilon_0=$ 20 GeV/fm$^3$ for a fixed
mass-shift. This is the reason for the transverse momentum spectrum for $\epsilon_0=$ 20 GeV/fm$^3$ is more affected than for $\epsilon_0=$ 40 GeV/fm$^3$.

\subsection{Effect of in-medium mass-shift on the elliptic flow}
\begin{figure}[htbp]
\includegraphics[scale=0.51]{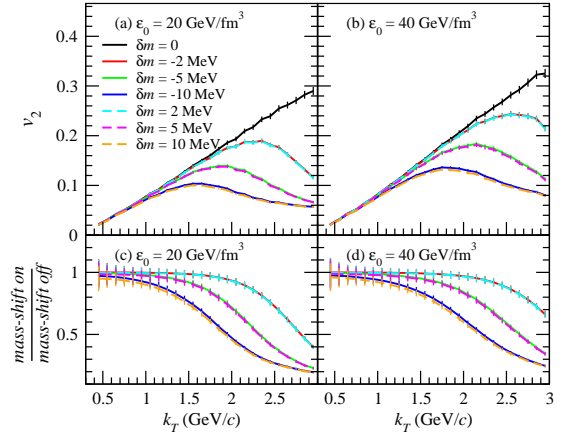}
\vspace*{0mm}
\caption{(Color online) The $v_2$ of $\phi$ meson as a function of $k_T$ with various mass-shift (top panel)
 and the ratio of $v_2$ of $\phi$ meson with mass-shift to which without mass-shift
(bottom panel) for the initial conditions $\epsilon_0=$ 20 and 40 GeV/fm$^3$ and $R_x=3$ fm, $R_y=4$ fm. }
\label{pv2}
\end{figure}

In Fig. \ref{pv2}, we show the $v_2$ of $\phi$ meson as a function of $k_T$ with various mass-shift
and the ratio of $v_2$ with mass-shift to which without mass-shift for the initial conditions
$\epsilon_0=$ 20 and 40 GeV/fm$^3$ and $R_x=3$ fm, $R_y=4$ fm. The $v_2$ of $\phi$ meson for
$\epsilon_0=$ 40 GeV/fm$^3$ is a little greater than for $\epsilon_0=$ 20 GeV/fm$^3$ when there is
no mass-shift ($\delta$$m$ = 0). The $v_2$ is suppressed by a small mass-shift in large transverse momentum
regions, and this effect increases with the increasing mass-shift. The suppression effect for
$\epsilon_0=$ 20 GeV/fm$^3$ is a little stronger than for $\epsilon_0=$ 40 GeV/fm$^3$ (see Fig. \ref{pv2} (c) and (d)).

To analyse the effect of mass-shift on $v_2$, we rewrite the Eq. (\ref{SP2}) as
\begin{eqnarray}\label{SP3}
&&N_*(\mk)\!=\!N_1(\mk)+N_2(\mk),\nonumber\\
&&N_1(\mk)\!=\!\int \frac{g_i}{(2\pi)^3}d^4\sigma_{\mu}(r)k^\mu\, \!n'_{\mk'},\nonumber\\
&&N_2(\mk)\!=\!\int \frac{g_i}{(2\pi)^3}d^4\sigma_{\mu}(r)k^\mu\,F_2.
\end{eqnarray}

\begin{figure*}[htbp]
\includegraphics[scale=1]{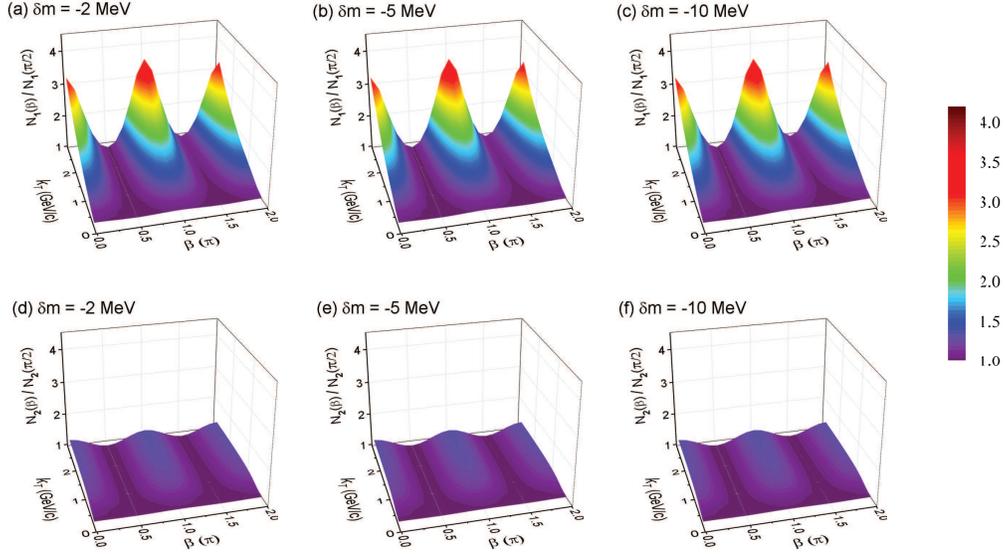}
\vspace*{-0.5mm}
\caption{(Color online) The ratio of $N_1$($\beta$) to $N_1$($\pi$/2) (top panel) and the ratio
of $N_2$($\beta$) to $N_2$($\pi$/2) (bottom panel) in $k_T$-$\beta$ plane for
the initial condition $\epsilon_0=$ 20 GeV/fm$^3$ and $R_x=3$ fm, $R_y=4$ fm. }
\label{20gn}
\end{figure*}
\begin{figure*}[htbp]
\includegraphics[scale=1]{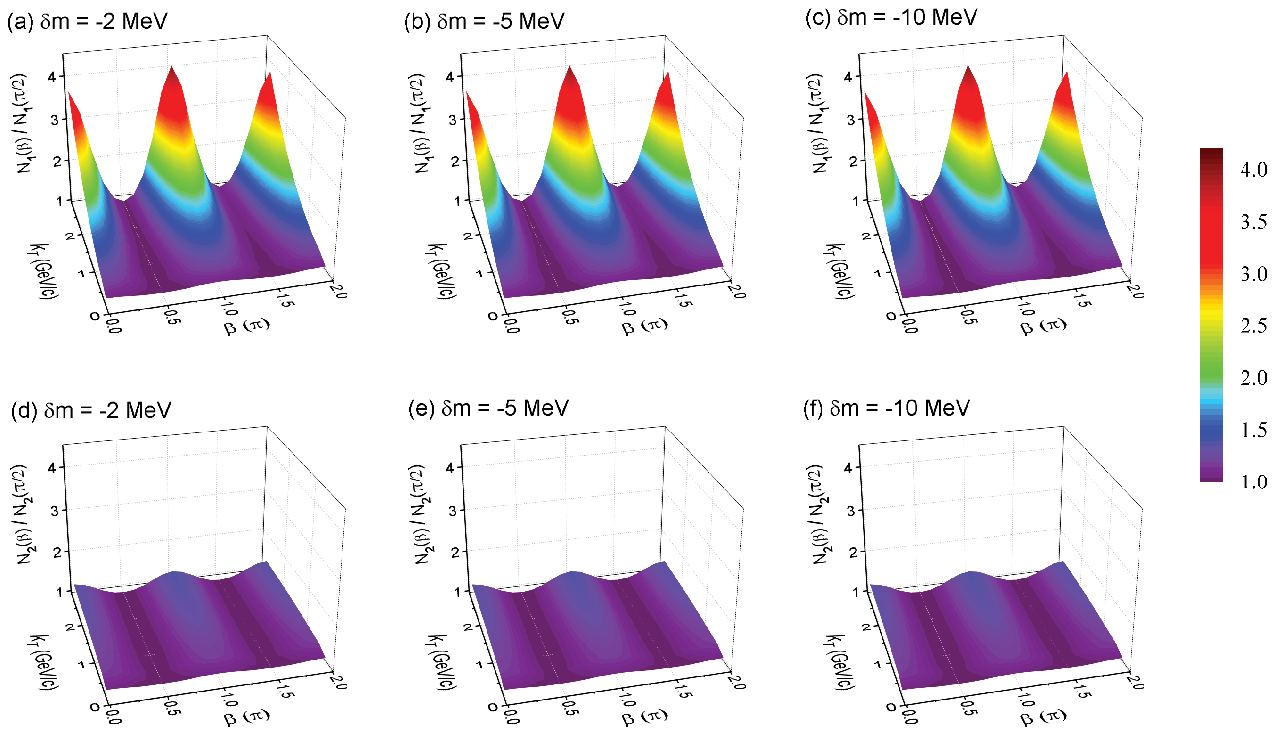}
\vspace*{-0.5mm}
\caption{(Color online) The ratio of $N_1$($\beta$) to $N_1$($\pi$/2) (top panel) and the ratio
of $N_2$($\beta$) to $N_2$($\pi$/2) (bottom panel) in $k_T$-$\beta$ plane for the
initial condition $\epsilon_0=$ 40 GeV/fm$^3$ and $R_x=3$ fm, $R_y=4$ fm. }
\label{40gn}
\end{figure*}

In Fig. \ref{20gn}, we show the ratio of $N_1$($\beta$) to $N_1$($\pi$/2) and the
ratio of $N_2$($\beta$) to $N_2$($\pi$/2) for the initial condition
$\epsilon_0=$ 20 GeV/fm$^3$ and $R_x=3$ fm, $R_y=4$ fm. Where $\beta$ is the
azimuthal angle of $\phi$ meson
\begin{equation}
\cos\beta=k_x/|\mk_T|,~~~~
\left(|\mk_T|=\sqrt{k_x^2+k_y^2}\,\right).
\end{equation}
The ratio of $N_1$($\beta$) to $N_1$($\pi$/2) and the ratio of $N_2$($\beta$)
to $N_2$($\pi$/2) can express the levels of the elliptical anisotropy of
$N_1(\mk)$ and $N_2(\mk)$ in Eq. (\ref{SP3}), respectively. The ratio of $N_1$($\beta$) to $N_1$($\pi$/2)
is almost equal to the ratio of $N_2$($\beta$) to $N_2$($\pi$/2) in low transverse momentum region,
so the mass-shift does not affect the $v_2$ in this region. However, the elliptical anisotropy of
$N_1(\mk)$ is much greater than $N_2(\mk)$ in large transverse momentum region, so the $v_2$ is suppressed by mass-shift in large transverse momentum region. Although the elliptical anisotropy of $N_2(\mk)$ is
almost the same for different $\delta$$m$, but the ratio of $\overline{F_2}$ to $\overline{n'_{\mk'}}$
for large $\delta$$m$ is greater than
for small $\delta$$m$, so the $v_2$ is more suppressed by a large mass-shift.

In Fig. \ref{40gn}, we show the ratio of $N_1$($\beta$) to $N_1$($\pi$/2) and the
ratio of $N_2$($\beta$) to $N_2$($\pi$/2) for the initial condition
$\epsilon_0=$ 40 GeV/fm$^3$. The elliptical anisotropy of $N_1(\mk)$ for $\epsilon_0=$ 40 GeV/fm$^3$
is a little greater than for $\epsilon_0=$ 20 GeV/fm$^3$ in large transverse momentum region, and thus leads to a more greater $v_2$ of $\phi$ meson without mass-shift for $\epsilon_0=$ 40 GeV/fm$^3$ (see Fig. \ref{pv2}). The elliptical
anisotropy of $N_2(\mk)$ for $\epsilon_0=$ 20 GeV/fm$^3$ is almost the same as for $\epsilon_0=$ 40 GeV/fm$^3$,
but the ratio of $\overline{F_2}$ to $\overline{n'_{\mk'}}$ for $\epsilon_0=$ 40 GeV/fm$^3$ is smaller than
for $\epsilon_0=$ 20 GeV/fm$^3$. And thus the $v_2$ for $\epsilon_0=$ 40 GeV/fm$^3$ is less suppressed than for $\epsilon_0=$ 20 GeV/fm$^3$ for a fixed mass-shift.

\subsection{Parts of all $\phi$ meson have a mass-shift}
When parts of all $\phi$ meson have a mass-shift, the single particle momentum distribution becomes
\begin{eqnarray}\label{SPpart}
 N^p_{*}(\mk)\!&&=(1-p)\!\int \frac{g_i}{(2\pi)^3}d^4\sigma_{\mu}(r)k^\mu\,n^0_{\mk'}\nonumber\\
 &&+\,p\!\int \frac{g_i}{(2\pi)^3}d^4\sigma_{\mu}(r)k^\mu\, \! [F_1\,n'_{\mk'}+F_2],
\end{eqnarray}
where $p$ is the probability of $\phi$ meson with mass-shift.

\begin{figure}[htbp]
\includegraphics[scale=0.54]{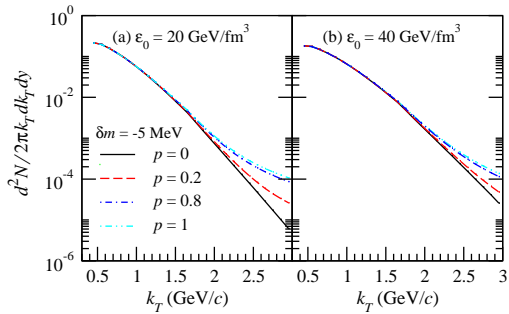}
\vspace*{-1mm}
\caption{(Color online) The normalized transverse momentum spectrum for parts of all $\phi$ meson
have a mass-shift, where the initial conditions are the same as in Fig. \ref{psp}.}
\label{psppart}
\end{figure}
\begin{figure}[htbp]
\includegraphics[scale=0.49]{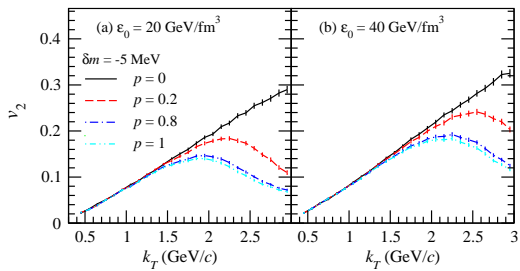}
\vspace*{-1mm}
\caption{(Color online) The $v_2$ for parts of all $\phi$ meson
have a mass-shift, where the initial conditions are the same as in Fig. \ref{pv2}.}
\label{pv2part}
\end{figure}

In Fig. \ref{psppart} and \ref{pv2part}, we show the normalized transverse momentum spectrum and
$v_2$ for parts of all $\phi$ meson with mass-shift. The effect of mass-shift on the transverse
momentum spectrum and $v_2$ decreases with the decreasing probability of $\phi$ meson
with mass-shift.

The above results indicate that the different mass-shift leads to different
transverse momentum spectrum and $v_2$.
The ranges of mass-shift of $\phi$ meson predicted by theory can be restricted by
comparing the simulated transverse momentum spectrum and $v_2$ to the experimental
data in high-energy heavy-ion collisions.

\section{Summary}
In this paper we studied the effect of in-medium mass-shift on the transverse-momentum spectrum
and $v_2$ of $\phi$ meson. The mass-shift leads to an
increase in the yield of $\phi$ meson in large transverse momentum region and
thus reduce the slope of the transverse-momentum spectrum,
and the elliptic flow $v_2$ of $\phi$ meson is suppressed by the mass-shift in large transverse momentum region. The effects of
mass-shift on the transverse-momentum spectrum and $v_2$ increase with the increasing mass-shift.
The effects also decrease with the increasing expanding velocity of the source. We further studied the effect of mass-shift on the transverse-momentum spectrum and $v_2$ for parts of all $\phi$ meson
with mass-shift, and the effect of mass-shift on the transverse-momentum spectrum and $v_2$
decreases with the decreasing probability of $\phi$ meson with mass-shift. Our study may
provide a way to restrict the ranges of mass-shift of $\phi$ meson by comparing the
simulated transverse-momentum spectrum and $v_2$ to the experimental data in high-energy heavy-ion collisions.

\begin{acknowledgments}
This research was supported by the National Natural Science Foundation
of China under Grant No. 11647166, 11747155, the Natural Science Foundation of Inner Mongolia
under Grant No. 2017BS0104, Changzhou Science and Technology Bureau CJ20180054, and the Foundation of Jiangsu University of Technology
under Grant No. KYY17028, KYY18048.
\end{acknowledgments}

\end{document}